\documentclass[preprint,12pt]{elsarticle}
\usepackage{graphicx,amsmath,latexsym}

\def\bea{\begin{eqnarray}}
\def\eea{\end{eqnarray}}
\def\be{\begin{equation}}
\def\ee{\end{equation}}
\newcommand{\ub}[1]{\underline{#1}}

\journal{Annals of Physics}

\begin{document}

\begin{frontmatter}

\title{On the nonperturbative solution
of Pauli--Villars-regulated light-front QED: \\
A comparison of the sector-dependent and standard parameterizations
}

\author{Sophia S. Chabysheva}
\author{John R. Hiller\corref{cor1}}
\ead{jhiller@d.umn.edu}
\cortext[cor1]{Corresponding author}
\address{Department of Physics \\
University of Minnesota-Duluth \\
Duluth, Minnesota 55812}

\begin{abstract}
We consider quantum electrodynamics quantized on the light front
in Feynman gauge and regulated in the ultraviolet by the inclusion
of massive, negative-metric Pauli--Villars (PV) particles in the
Lagrangian.  The eigenstate of the electron is approximated by
a Fock-state expansion truncated to include one photon.  The Fock-state
wave functions are computed from the fundamental Hamiltonian 
eigenvalue problem and used to calculate the anomalous magnetic
moment, as a point of comparison.  Two approaches are considered: a 
sector-dependent parameterization, where the bare parameters
of the Lagrangian are allowed to depend on the Fock sectors
between which the particular Hamiltonian term acts, and the
standard choice, where the bare parameters are the
same for all sectors.  Both methods are shown to require some
care with respect to ultraviolet divergences; neither method
can allow all PV masses to be taken to infinity.  In addition,
the sector-dependent approach suffers from an infrared divergence
that requires a nonzero photon mass; due to complications
associated with this divergence, the standard parameterization
is to be preferred.  We also show that the
self-energy effects obtained from a two-photon truncation
are enough to bring the standard-parameterization result for the
anomalous moment into agreement with experiment within numerical
errors.  This continues the development of a method for the
nonperturbative solution of strongly coupled theories, in
particular quantum chromodynamics.
\end{abstract}
\begin{keyword}
light-cone quantization \sep Pauli--Villars regularization \sep
quantum electrodynamics \sep anomalous magnetic moment
\PACS 12.38.Lg \sep 11.15.Tk \sep 11.10.Gh \sep 11.10.Ef
\end{keyword}

\end{frontmatter}

%%%%%%%%%%%%%%%%%%%%%%
\section{Introduction}
\label{sec:Introduction}
%%%%%%%%%%%%%%%%%%%%%%

The nonperturbative solution of quantum field theories has
proven to be a difficult task.  The method that has had the
most success to date, lattice gauge theory~\cite{lattice},
has attained this success only after a long period of development,
with a great number of technical innovations along the way.
What is more, the level of success that can be achieved is
inherently limited by the lack of direct contact with wave
functions for bound-state constituents and by the Euclidean
formulation.  The limitation of Euclidean formulation is
shared by the method of Dyson--Schwinger equations~\cite{DSE},
which are coupled equations for the $n$-point Euclidean Green's
functions where bound states appear as poles in the propagators.
Solution of the infinite system requires truncation and a model
for the highest $n$-point functions. 

In order to calculate wave functions directly in a Minkowskian
formulation, there has been an effort for a number of years to
develop a Hamiltonian approach in light-cone quantization~\cite{DLCQreview}.
Although the ultimate objective is to be able to solve for the bound states
of QCD, most of the development has been in QED and simpler
theories, such as Yukawa theory and $\phi^4$ theory.  Many two-dimensional
theories have been solved; however, success with four-dimensional
theories has been limited by the need for a consistent regularization
and renormalization scheme and by the large size of the numerical
calculation.  There is also a related light-front lattice Hamiltonian
formulation, known as the transverse lattice method~\cite{TransLattice},
which we do not consider here.

A regularization scheme that has been useful in doing perturbative
light-cone calculations is the alternate denominator method 
of Brodsky, Roskies, and Suaya~\cite{BRS}.  Because of its
success, one would naturally consider extending this approach
to nonperturbative calculations.  Unfortunately, application
of the method requires explicit identification of light-cone
energy denominators, which are only implicit in the coupled
equations of the nonperturbative mass eigenvalue problem.
One could instead use the alternate denominator method to
construct counterterms, and then incorporate the counterterms
in the Hamiltonian for the nonperturbative calculation.  
However, these counterterms would be limited to a particular
order in the coupling while the nonperturbative problem
sums a partial set of contributions to all orders.  Instead,
one needs a method that generates counterterms to all orders,
but no more than what is needed.

A method for regularization that has proven quite useful is 
Pauli--Villars (PV) regularization~\cite{PauliVillars}, which
was developed and tested in a series of calculations~\cite{bhm1,bhm2,%
YukawaDLCQ,ExactSolns,YukawaOneBoson,OnePhotonQED,YukawaTwoBoson,%
ChiralLimit}.  The key idea is to include enough PV fields in
the Lagrangian to regulate the theory perturbatively
and, where possible, maintain symmetries.\footnote{The introduction
of PV partners to the fields of a theory has recently
been used to define extensions of the Standard Model
that offer a solution to the hierarchy problem~\protect\cite{Lebed}.}
The derived
light-front Hamiltonian then defines the nonperturbative bound-state
problem.  As for the eigenstate, it is approximated by a truncated Fock-state
expansion.  Then the mass eigenvalue problem leads to coupled integral
equations for the Fock-state wave functions.  The PV particles
appear in the Fock states and, through chosen negative metrics,
bring about the subtractions necessary to regulate the integral
equations.  A possible formulation for QCD along these lines
has been given by Paston {\em et al}.~\cite{Paston}.

Two methods of parameterization are in use.  One is the
standard choice,
where the bare parameters are those of the regulated Lagrangian.
The other is a sector-dependent parameterization~\cite{SectorDependent},
where the bare parameters are allowed to depend on the Fock sector(s)
on which the Hamiltonian acts.  In either case, the parameters
are fixed by constraints from observables and from symmetry restorations.
The standard scheme has been used extensively in studies 
of PV regularization~\cite{bhm1,bhm2,%
YukawaDLCQ,ExactSolns,YukawaOneBoson,OnePhotonQED,YukawaTwoBoson,%
ChiralLimit}.  The sector-dependent scheme was first systematically
applied to QED by Hiller and Brodsky~\cite{hb}, though they did
not consider a sector-dependent vertex mass.  More recently,
it was investigated by Karmanov, Mathiot, and Smirnov~\cite{Karmanov}.
In order to better understand how to proceed with the large-scale
numerical calculations that need to be done, a comparison of these
approaches needs to be made.

Both parameterizations require some care, particularly with
respect to uncanceled divergences.  In the standard method, 
an uncanceled divergence appears in the following way~\cite{OnePhotonQED}.  
The results for a generic physical quantity, such as the 
electron's anomalous moment, will be of the form
\be
\lim_{\mu_{\rm PV}\rightarrow \infty }
g^2\frac{a_1 \,[+a_2 g^2 \ln\mu_{\rm PV}+\cdots]}
                  {1+b_1g^2+b_2g^2\ln\mu_{\rm PV}+\cdots}
= \left\{\begin{array}{ll} 0, & \mbox{with truncation} \\
                                   \mbox{finite}, & \mbox{without truncation},
                                     \end{array} \right.
\ee
where $\mu_{\rm PV}$ is a PV mass scale and the contents of
the square brackets are absent in the case of truncation.
When the limit $\mu_{\rm PV}\rightarrow\infty$ is taken, the
result is either zero or a finite value.  As the PV mass
$\mu_{\rm PV}$ is increased, the divergence in the denominator
is not canceled by a divergence in the numerator, if the
truncation is made.  Therefore, any potentially meaningful
calculation must be done at finite $\mu_{\rm PV}$, in a
range of scales where the errors due to truncation and
PV inclusion are minimized.

The sector-dependent method is also limited to finite
PV mass scales by an uncanceled divergence.  
As shown in \cite{hb}, the bare coupling
can become imaginary if the PV masses become too large.
Also, as we will show below, the probabilities for individual
Fock sectors are driven outside the range of 0 to 1.
The authors of \cite{Karmanov} do not consider these
limitations and calculate quantities that do not depend 
directly on this bare coupling;
they then have a result at infinite PV masses, but the
underlying theory and its wave functions are ill defined.

Instead, the calculation should be done at finite PV masses.
Here we do this, for the anomalous magnetic moment of the
electron computed in the one-photon truncation of the
dressed-electron state, and compare the results of
the standard~\cite{ChiralLimit} and 
sector-dependent~\cite{Karmanov} approaches.
In this particular context, the dressed electron can be viewed
as a bound state of bare constituents, electron and photon, in QED.
The particular truncation is chosen because the eigenvalue
problem can be solved analytically and the comparison
is not obscured by numerical artifacts.  The calculation
of the anomalous moment is not intended as competitive
with perturbation theory~\cite{Kinoshita}; the weak
coupling of QED makes nonperturbative calculations much
less accurate.  However, the calculation does stand as
a test of a method that can be applied where perturbation
theory is useless.

Convergence with respect to the Fock-state expansion has
been checked in Yukawa theory~\cite{YukawaDLCQ}.  There
a full discrete light-cone quantization (DLCQ)~\cite{PauliBrodsky}
calculation was possible.  The DLCQ approach does truncate the
Fock-state expansion but only to the extent required by the 
numerical resolution; more constituents are allowed as the
discrete momentum fraction is reduced.  Explicit truncations
to fewer constituents can be made and the results compared.
For Yukawa theory, the comparison showed that the Fock-state expansion
converges quickly.

The analogous results for a cubic scalar theory~\cite{KarmanovMaris} 
do not show such rapid convergence.  However, this is because
the spectrum of the theory is unbounded from below~\cite{Baym},
and the lowest eigenstates are dominated by Fock states with large
numbers of constituents~\cite{ZeroModes}.  In other words, for
cubic scalar theories the highest Fock states make the largest
contributions.  Yukawa theory and
QED do not suffer this fate because of the Pauli principle.

In order to have a simple vacuum and a well-defined
Fock-state expansion, we use light-cone
coordinates~\cite{Dirac,DLCQreview} $\ub{x}=(x^-=t-z,\vec x_\perp=(x,y))$
and light-cone momenta $\ub{p}=(p^+=p^0+p^z,\vec p_\perp)$.
The light-cone energy $p^-=p^0-p^z$ is conjugate to
the light-cone time $x^+=t+z$.  
Stationary states are obtained as eigenstates of the
light-cone Hamiltonian~\cite{DLCQreview}
${\cal P}^-$, which defines the
mass eigenvalue problem ${\cal P}^-|P\rangle=\frac{M^2}{P^+}|P\rangle$,
in a frame where the total transverse momentum $\vec P_\perp$
is zero.  Further details of the coordinate choice can be 
found in \cite{ChiralLimit} or \cite{DLCQreview}.

To further develop the method in the context of a gauge theory,
we consider QED.
To regulate QED, we use one PV electron and two PV photons.
The second PV photon is needed to restore chiral symmetry
in the massless-electron limit~\cite{ChiralLimit}.  The
couplings and metrics of the PV fields are 
adjusted to accomplish the ultraviolet
regularization and the finite chiral-symmetry correction.
They also provide for cancellation of instantaneous fermion
interactions and allow the constraint equation for the 
nondynamical fermion fields to be solved exactly without
use of light-cone gauge~\cite{ChiralLimit}.

In Sec.~2 of \cite{OnePhotonQED}, there is some discussion
of the use of three PV fermions to regulate QED.  However, this
was included only to show that naive PV regularizations in
light-cone gauge do not necessarily work.  A regularization
in light-cone gauge that does work is given in Sec.~4 of
that paper, but this requires additional regulators that
complicate the theory.  The simplest regularization, which
uses one PV fermion and one PV boson, is found for Feynman
gauge and discussed in Sec.~3.  There the chiral limit is
correct, but only because the PV electron mass is taken
to infinity.  If the PV electron mass is kept finite,
regularization by one PV photon is insufficient for both
standard and sector-dependent parameterizations.

Also, the result in \cite{OnePhotonQED} for the anomalous
moment of the electron differs from the Schwinger result~\cite{Schwinger}.
The source of the difference is not the uncanceled divergence.  
Instead, it is contributions from all higher orders in the
coupling constant, consistent
with the Fock-space truncation to one electron and
one photon.
The effect of the uncanceled divergence,
which appears in the denominators of Fock-state probabilities,
is logarithmic and quite mild, and, in any case, can only
make the estimate for the anomalous moment smaller than
the Schwinger result.  The estimate obtained is instead
larger, due to sensitivity to the constituent (bare) electron
and photon masses and
the nonperturbative shift of the constituent electron mass.
Here we correct this shift by including the self-energy contribution
from the one-electron/two-photon sector and find very good
agreement with experiment.  

The sector-dependent approach,
as used in \cite{Karmanov}, does no better than lowest-order
perturbation theory, which is quite strange, since a
nonperturbative calculation should include some physics
to all orders.  In addition, the calculation in \cite{Karmanov}
suffers from uncanceled divergences that appear in the
numerators of Fock-state probabilities, instead of the
denominators, and that push these probabilities outside
the physical range of $[0,1]$.

With this as introduction, we continue in Sec.~\ref{sec:onephoton}
with a comparison of the
standard and sector-dependent parameterizations in a one-photon
truncation of the Fock expansion for the dressed electron.
In Sec.~\ref{sec:selfenergy}, the standard-parameterization
results are extended to include self-energy
contributions that correspond to intermediate states with
two photons.  Section~\ref{sec:summary} contains a 
summary of the comparison and some general discussion
of applications.  \ref{sec:FeynmanQED} provides specifics about 
light-front QED in Feynman gauge, with further details
available in \cite{OnePhotonQED} and \cite{ChiralLimit}.
\ref{sec:SEsolution} gives details of the solution when self-energy
contributions are included.

%%%%%%%%%%%%%%%%%%%%%%%
\section{One-Photon Truncation}
\label{sec:onephoton}
%%%%%%%%%%%%%%%%%%%%%%

We work in Feynman-gauge QED quantized on the light-front and regulated
with one PV electron and two PV photons.  Details are given in
\ref{sec:FeynmanQED}.

\subsection{Fock-State Expansion}

We seek the solution for the dressed-electron eigenstate truncated
to include only the bare-electron and one-electron/one-photon 
Fock sectors.  The Fock-state expansion for the eigenstate with
total $J_z=\pm\frac12$ is then
\be \label{eq:FockExpansion}
|\psi^\pm(\ub{P})\rangle=\sum_i z_i b_{i\pm}^\dagger(\ub{P})|0\rangle
  +\sum_{ijs\mu}\int d\ub{k} C_{ijs}^{\mu\pm}(\ub{k})b_{is}^\dagger(\ub{P}-\ub{k})
                                       a_{j\mu}^\dagger(\ub{k})|0\rangle.
\ee
The amplitudes $z_i$ and wave functions $C_{ijs}^{\mu\pm}$
define this state.  

The interpretation of the expansion
requires projection onto a physical subspace.
We apply the same approach as was used in Yukawa
theory~\cite{YukawaOneBoson}, where a projection onto 
the physical subspace is accomplished
by expressing Fock states in terms of positively normed
creation operators and null combinations.  Here, the
positive-norm operators are
$a_{0\mu}^\dagger$, $a_{2\mu}^\dagger$, and 
$b_{0s}^\dagger$, and the null combinations are
$a_\mu^\dagger=\sum_i \sqrt{\xi_i}a_{i\mu}^\dagger$ 
and $b_s^\dagger=b_{0s}^\dagger+b_{1s}^\dagger$.
The $b_s^\dagger$ particles are annihilated by the
generalized electromagnetic current $\bar\psi\gamma^\mu\psi$; thus,
$b_s^\dagger$ creates unphysical contributions to be dropped,
and, by analogy, we also drop contributions created by $a_\mu^\dagger$.
The projected dressed-fermion state is
\bea \label{eq:projected}
\lefteqn{|\psi_{\rm phys}^\pm(\ub{P})\rangle=\sum_i (-1)^i z_i
                                          b_{0\pm}^\dagger(\ub{P})|0\rangle} && \\
  &&+\sum_{s\mu}\int d\ub{k} \sum_{i=0}^1\sum_{j=0,2}\sqrt{\xi_j}
        \sum_{k=j/2}^{j/2+1} \frac{(-1)^{i+k}}{\sqrt{\xi_k}}
                        C_{iks}^{\mu\pm}(\ub{k})
               b_{0s}^\dagger(\ub{P}-\ub{k})
                                       a_{j\mu}^\dagger(\ub{k})|0\rangle . \nonumber
\eea
The normalization condition
\be  \label{eq:norm}
\langle\psi_{\rm phys}^{\sigma'}(\ub{P}')|\psi_{\rm phys}^\sigma(\ub{P})\rangle
                  =\delta(\ub{P}'-\ub{P})\delta_{\sigma'\sigma}
\ee
is then applied.

The projection is critical, not only for the removal of negatively
normed contributions which could make probabilistic interpretations
difficult, but also for the regularization of expectation values.
In particular, a calculation of the expectation value of the
number of photons in the dressed-electron state will yield infinity
for the unprojected Fock expansion, because the spin-flip wave 
function $C_{ij\mp}^{\mu\pm}(\ub{k})$ falls off inversely with $k_\perp$.
The projection introduces subtractions that remove this divergent
behavior.  In \cite{Karmanov}, such a projection is not made and 
instead the eigenstate is to be interpreted only as part of
a larger process; the associated Fock-state wave functions
have no direct utility.

\subsection{Coupled Equations}

Our Fock-space-truncated eigenstate has to satisfy the mass-squared
eigenvalue problem ${\cal P}^-|\psi\rangle=\frac{M^2}{P^+}|\psi\rangle$.
Projections onto each Fock sector yield coupled equations for the
one-photon amplitudes $z_i$ and one-electron/one-photon wave
functions $C_{ijs}^{\mu\pm}$.
These coupled equations are, with $y=k^+/P^+$,
\bea \label{eq:firstcoupledequation}
[M^2-m_i^2]z_i & = & \int (P^+)^2 dy d^2k_\perp 
     \sum_{j,l,\mu}\sqrt{\xi_l}(-1)^{j+l}\epsilon^\mu \\
&& \times
  \left[V_{ji\pm}^{\mu*}(\ub{P}-\ub{k},\ub{P})C^{\mu\pm}_{jl\pm}(\ub{k}) 
        +U_{ji\pm}^{\mu*}(\ub{P}-\ub{k},\ub{P}) C^{\mu\pm}_{jl\mp}(\ub{k})\right] , \nonumber
\eea
and
\be \label{eq:TwoBodyEqns}
\left[M^2 - \frac{m_j^2 + k_\perp^2}{1-y} - \frac{\mu_l^2 + k_\perp^2}{y}\right]
  C^{\mu\pm}_{jl\pm}(\ub{k})
    =\sqrt{\xi_l}\sum_{i'} (-1)^{i'} z_{i'} P^+ V_{ji'\pm}^\mu(\ub{P}-\ub{k},\ub{P}),
\ee
\be \label{eq:TwoBodyEqns2}
\left[M^2 - \frac{m_j^2 + k_\perp^2}{1-y} - \frac{\mu_l^2 + k_\perp^2}{y}\right]
C^{\mu\pm}_{jl\mp}(\ub{k}) 
    = \sqrt{\xi_l}\sum_{i'} (-1)^{i'} z_{i'} P^+ U_{ji'\pm}^\mu(\ub{P}-\ub{k},\ub{P}).
\ee
The indices are arranged such that an index of $i$ corresponds to the 
one-electron sector and $j$ to the one-electron/one-photon sector.  
Therefore, in the sector-dependent approach, a mass $m_i$ in a vertex
function is assigned the bare mass, and $m_j$ is the physical
mass.  In the standard parameterization, all are bare masses.

The coupled equations can be solved analytically~\cite{OnePhotonQED}. 
The wave functions $C_{ils}^{\mu\pm}$ are
\begin{eqnarray} \label{eq:wavefn1}
C^{\mu\pm}_{il\pm}(\ub{k}) &=& \sqrt{\xi_l}
  \frac{\sum_j (-1)^j z_j P^+ V_{ij\pm}^\mu(\ub{P}-\ub{k},\ub{P})}
    {M^2 - \frac{m_i^2 + k_\perp^2}{1-y} - \frac{\mu_l^2 + k_\perp^2}{y}} , \\
\label{eq:wavefn2}
C^{\mu\pm}_{il\mp}(\ub{k}) &=& \sqrt{\xi_l}
\frac{\sum_j (-1)^j z_j P^+ U_{ij\pm}^\mu(\ub{P}-\ub{k},\ub{P})}
     {M^2 - \frac{m_i^2 + k_\perp^2}{1-y} - \frac{\mu_l^2 + k_\perp^2}{y}} ,
\end{eqnarray}
and the amplitudes satisfy
\be \label{eq:FeynEigen}
(M^2-m_i^2)z_i =
      2e_0^2\sum_{i'} (-1)^{i'}z_{i'}\left[\bar{J}+m_im_{i'} \bar{I}_0
  -2(m_i+m_{i'}) \bar{I}_1 \right],
\ee
with~\cite{OnePhotonQED}
\begin{eqnarray} \label{eq:In}
\bar{I}_n(M^2)&=&\int\frac{dy dk_\perp^2}{16\pi^2}
   \sum_{jl}\frac{(-1)^{j+l}\xi_l}{M^2-\frac{m_j^2+k_\perp^2}{1-y}
                                   -\frac{\mu_l^2+k_\perp^2}{y}}
   \frac{m_j^n}{y(1-y)^n}\,, \\
\bar{J}(M^2)&=&\int\frac{dy dk_\perp^2}{16\pi^2}  \label{eq:J}
   \sum_{jl}\frac{(-1)^{j+l}\xi_l}{M^2-\frac{m_j^2+k_\perp^2}{1-y}
                                   -\frac{\mu_l^2+k_\perp^2}{y}}
   \frac{m_j^2+k_\perp^2}{y(1-y)^2} .
\end{eqnarray}
Each integral can be computed analytically.
For the sector-dependent parameterization,
the convention of $m_i$ being a bare mass and $m_j$ a physical mass
has been maintained, with the extension that $m_{i'}$ is also
a bare mass. The integrals $\bar{I}_n$ and $\bar{J}$ then depend
only on the physical mass, and (\ref{eq:FeynEigen}) is equivalent to perturbation
theory.  For the standard parameterization, $\bar{I}_n$ and $\bar{J}$
depend on the bare mass, and (\ref{eq:FeynEigen}) is nonperturbative.

With use of the identity~\cite{ChiralLimit} $\bar{J}=M^2 \bar{I}_0$,
$\bar J$ can be eliminated from the eigenvalue problem.
This allows the solution to the eigenvalue problem
(\ref{eq:FeynEigen}) to take the simple form~\cite{OnePhotonQED}
\begin{equation} \label{eq:OneBosonEigenvalueProb}
\alpha_{0\pm}=\frac{(M\pm m_0)(M\pm m_1)}{8\pi (m_1-m_0)(2 \bar{I}_1\pm M\bar{I}_0)} , \;\;
z_1=\frac{M \pm m_0}{M \pm m_1}z_0 ,
\end{equation}
with $\alpha_0=e_0^2/4\pi$ and $z_0$ determined by the normalization (\ref{eq:norm}),
which reduces to
\bea \label{eq:reducednorm}
1&=&(z_0-z_1)^2 \\
&& +\frac{\alpha_0}{2\pi}\int y dy dk_\perp^2\sum_{l,l'}(-1)^{l+l'}z_l z_{l'}
  \sum_{i'i}(-1)^{i'+i}\sum_{j=0,2}\xi_j \nonumber \\
  && \sum_{k'=j/2}^{j/2+1}\sum_{k=j/2}^{j/2+1}
     \frac{(-1)^{k'+k}}{[ym_{i'}^2+(1-y)\mu_{k'}^2+k_\perp^2-m_e^2y(1-y)]} \nonumber \\
 && \times
    \frac{m_{i'} m_i-(m_i+m_{i'})(m_l+m_{l'})(1-y)+m_lm_{l'}(1-y)^2+k_\perp^2}
  {[ym_i^2+(1-y)\mu_k^2+k_\perp^2-m_e^2y(1-y)]} .
  \nonumber
\eea

\subsection{Discussion of the Solution}

Since the PV fermion is needed only to regulate the integral $\bar J$, which
has been eliminated from the calculation, we can safely take the 
$m_1\rightarrow\infty$ limit, to simplify the remaining steps.  In this
limit, we have $z_1=0$, $m_1z_1\rightarrow\pm(M-m_0)z_0$, and
\be  \label{eq:alpha0}
\alpha_{0\pm}=\pm\frac{M(M\pm m_0)}{8\pi (2 \bar{I}_1\pm M\bar{I}_0)}.
\ee
Also, the third constraint (\ref{eq:3rdconstraint}) is automatically
satisfied, and the second PV photon flavor can be discarded.

For the standard parameterization, we cannot solve explicitly for
$m_0$, because $\bar I_0$ and $\bar I_1$ are functions of
$m_0$.  However, $\alpha_{0\pm}$ is just $\alpha_\pm$
and the value of $m_0$ is
determined by requiring $\alpha_\pm$ to be equal to the
physical value of $\alpha$.  This defines a nonlinear equation
for $m_0$, due to the dependence of $\bar I_n$ on $m_0$.
For small values of the PV masses
there may be no such solution; however, for reasonable values
we do find at least one solution for each branch.
The plot in Fig.~1 of \cite{ChiralLimit} shows $\alpha_\pm/\alpha$
as functions of $m_0$.  The $\alpha_-$ branch is the
physical choice, because the no-interaction limit ($\alpha_-=0$)
corresponds to the
bare mass $m_0$ becoming equal to the physical electron
mass, $M=m_e$. This is consistent with the sector-dependent case,
where the lower sign is also chosen.
 
If the PV electron has a sufficiently large mass,
the value of $m_0$ that yields $\alpha_-=\alpha$ is
less than $m_e$.  In this case, the integrals $\bar I_n$
and $\bar J$ contain poles for $j=l=0$ and are defined
by a principal-value prescription~\cite{OnePhotonQED}.
For terms in the normalization sum (\ref{eq:reducednorm})
with $i=k=0$ or $i'=k'=0$, there are
simple poles, again defined by a principal-value prescription.  For the terms
where all four of these indices are zero, there is a double pole,
defined by the prescription~\cite{OnePhotonQED}
\begin{equation} \label{eq:prescrip}
\int dx \frac{f(x)}{(x-a)^2}\equiv  \lim_{\eta\rightarrow 0} \frac{1}{2\eta}
  \left[\mathcal{P}\int dx\frac{f(x)}{x-a-\eta}
          -\mathcal{P}\int dx \frac{f(x)}{x-a+\eta}\right].
\end{equation}

In the sector-dependent approach, $\bar I_1$ and $\bar I_0$ are independent
of $m_0$, and the solution for $\alpha_0$ in (\ref{eq:alpha0})
can be written as an explicit expression for $m_0$
\be
m_0=\mp M + 8\pi\frac{\alpha_{0\pm}}{M}(2 \bar{I}_1\pm M\bar{I}_0).
\ee
The lower sign is chosen, so that $m_0$ reduces to $M=m_e$ when
the coupling is zero.  This expression for $m_0$ is equivalent to
Eq.~(74) of \cite{Karmanov}, since the $m_1\rightarrow\infty$
limits of $\bar I_0$ and $\bar I_1$ are equivalent to $B$ and $A$
in \cite{Karmanov}.

There is, however, the sector dependence of the coupling.
For the coupling between the bare-electron
and one-electron/one-photon sectors, the bare coupling is
given by Eq.~(3.20) of \cite{hb}, which in our notation is
written $e_0=e/z_0$, where $z_0$ is the amplitude for the
bare-electron Fock state computed without projection onto
the physical subspace.  This expression arises from the
following considerations.  In general, the bare coupling
would be $e_0=Z_1e/\sqrt{Z_{2i}Z_{2f}Z_3}$; this includes
the truncation effect that splits the usual $Z_2$ into
a product of different $\sqrt{Z_2}$ from each fermion
leg~\cite{OSUQED}.  Since no fermion-antifermion loop
is included, we have $Z_3=1$.  Since only one photon
is included, there is no vertex correction and $Z_1=1$.
Also, only the fermion leg with no photon spectator
will be corrected by $\sqrt{Z_2}$ and therefore
$\sqrt{Z_{2i}Z_{2f}}=z_0$.

In the infinite-$m_1$ limit, the bare-electron amplitude
without projection is determined by the normalization
\be
1=z_0^2+e_0^2z_0^2\tilde J_2,
\ee
obtained as a limit of the unprojected form of (\ref{eq:reducednorm}),
with
\be \label{eq:tildeJ2}
\tilde J_2=\frac{1}{8\pi^2}\int y\, dy dk_\perp^2
  \sum_{k=0}^1(-1)^{k}
  \frac{(y^2+2y-2)m_e^2+k_\perp^2}
       {[k_\perp^2+(1-y)\mu_{k}^2+y^2m_e^2]^2}.
\ee
On replacement of $e_0$ by $e/z_0$, we can solve for $z_0$ as
\be
z_0=\sqrt{1-e^2\tilde J_2}
\ee
and also find
\be
e_0=e/\sqrt{1-e^2\tilde J_2}.
\ee
This result matches Eq.~(79) of \cite{Karmanov}, with
$\tilde J_2$ in our Eq.~(\ref{eq:tildeJ2}) agreeing with
Eq.~(77) in \cite{Karmanov}.  However, our interpretation is quite
different, in that we do not allow $e_0$ to become
imaginary, which limits $\tilde J_2$.
Since $\tilde J_2$ grows logarithmically with
$\mu_1$, as
\be \label{eq:J2log}
\tilde J_2\simeq \frac{1}{8\pi^2}\left(\ln\frac{\mu_1\mu_0^2}{m_e^3}+\frac98\right),
\ee
this last PV mass cannot be taken to infinity.
In \cite{Karmanov}, this behavior is disregarded, $\mu_1$ is
taken to infinity, and the eigenstate
is re-interpreted in the context of some larger process,
where $e_0$ is not directly referenced.  Unfortunately,
an imaginary $e_0$ makes the underlying theory unphysical,
with opposite charges repelling each other.  There would
also be strange results for probabilities when the
uncanceled divergence is ignored, so that
$\tilde J_2$ grows without limit.  The
probability for the one-electron sector is $z_0^2=1-e^2\tilde J_2<0$
and for the one-electron/one-photon sector, $e^2\tilde J_2>1$.

For the standard parameterization, there is also a limit
on $\mu_1$.  The projected normalization condition
(\ref{eq:reducednorm}) can be written as
\be
1=z_0^2+e^2z_0^2 J_2,
\ee
with
\bea \label{eq:J2}
J_2&=&\frac{1}{8\pi^2}\int y\, dy dk_\perp^2
   [m_0^2-4m_0m_e(1-y)+m_e^2(1-y)^2+k_\perp^2] \\
 && \times \left(\sum_{k=0}^1(-1)^k
  \frac{1}
       {[k_\perp^2+(1-y)\mu_{k}^2+ym_0^2-y(1-y)m_e^2]}\right)^2 .  \nonumber
\eea
Thus the bare amplitude is
\be \label{eq:z0SR}
z_0=1/\sqrt{1+e^2J_2},
\ee
which is driven to zero as $\mu_1\rightarrow\infty$ and causes
most expectation values also to go to zero.  This is again the
difficulty of uncanceled divergences~\cite{OnePhotonQED}.  Here
the denominator contains a term one order higher in $\alpha$
than the truncation allows in the numerator.  Any divergence
in this higher-order term is not canceled by a corresponding
divergence in the numerator, and the expectation value
becomes zero.

To obtain meaningful results for both standard and sector-dependent
approaches, one must balance minimization
of the truncation error against minimization of the errors
associated with having finite-mass, negative-metric PV
particles in the basis~\cite{OnePhotonQED}.  We look for
a range of PV masses where expectation values are approximately
constant, and find that this does occur for the 
anomalous moment.  Since $J_2$ and $\tilde J_2$ grow 
logarithmically with $\mu_1$, the dependence is very mild,
and $\mu_1$ can be taken quite large.

There is an additional complication of an infrared divergence
in the sector-dependent case.  The pole in the
wave function (\ref{eq:wavefn1})
moves to an endpoint when $\mu_0=0$ and the constituent
mass $m_0$ is set equal to the physical mass $m_e$.
This happens because the denominator of the wave function at zero
transverse momentum is proportional to $m_e^2-m_0^2/(1-y)-\mu_0^2/y$.
At the endpoint, a principal-value prescription cannot
be used, and the mass of the physical photon $\mu_0$
must be nonzero to remove this pole.
However, the anomalous moment of the electron is particularly
sensitive to this mass, and even a small
value for $\mu_0$ will cause a noticeable shift~\cite{hb}.
From another side, the integral $\tilde J_2$
is driven to negative values as $\mu_0$ goes to zero, 
as can be seen from (\ref{eq:J2log}), making
the probability of the two-particle sector negative and
therefore unphysical.  Thus, the $\mu_0\rightarrow0$ limit
is needed for an accurate result, but the limit cannot be
taken.
The standard parameterization does not
suffer from this infrared problem, because the bare mass
is never equal to the physical electron mass.

%%%%%%%%%%%%%%%%%%%%%%%%%%%%%%%%%%%%%%%%%%%%%%%%%%%%%%%%%
\subsection{Anomalous Magnetic Moment}
\label{sec:anomalousmoment} 

To compute the anomalous moment, we
start from the Brodsky--Drell formula~\cite{BrodskyDrell}
derived from the spin-flip matrix element of the electromagnetic current.
In the one-photon truncation, the formula reduces to
\bea
a_e&=&m_e\sum_{s\mu}\int d\ub{k}\epsilon^\mu \sum_{j=0,2}\xi_j
  \left(\sum_{i'=0}^1\sum_{k'=j/2}^{j/2+1}
    \frac{(-1)^{i'+k'}}{\sqrt{\xi_{k'}}}C_{i'k's}^{\mu+}(\ub{k})\right)^* \\
  && \times y\left(\frac{\partial}{\partial k_x}+i\frac{\partial}{\partial k_y}\right)
  \left(\sum_{i=0}^1\sum_{k=j/2}^{j/2+1}
    \frac{(-1)^{i+k}}{\sqrt{\xi_k}}C_{iks}^{\mu-}(\ub{k})\right). \nonumber
\eea
Given the explicit expressions (\ref{eq:wavefn1}) and (\ref{eq:wavefn2})
for the wave functions $C_{iks}^{\mu\pm}$, the expression for $a_e$ simplifies to
\bea \label{eq:OnePhotonae}
a_e&=&\frac{\alpha_0}{\pi}m_e\int y^2 (1-y) dy dk_\perp^2
\sum_{l,l'}(-1)^{l+l'}z_l z_{l'}m_l\sum_{j=0,2}\xi_j \\
  &&  \times\left(\sum_{i=0}^1\sum_{k=j/2}^{j/2+1} \frac{(-1)^{i+k}}{ym_i^2+(1-y)\mu_k^2+k_\perp^2-m_e^2y(1-y)}\right)^2 . \nonumber
\eea
In the sector-dependent case, $m_i$ is a physical mass but $m_l$ is a bare mass.
The double pole that occurs in the case of the standard parameterization is handled 
in the same way as for the normalization
integrals, as discussed above.  The integrals can be done analytically.

\begin{figure}[ht]
\vspace{0.2in}
\centerline{\includegraphics[width=15cm]{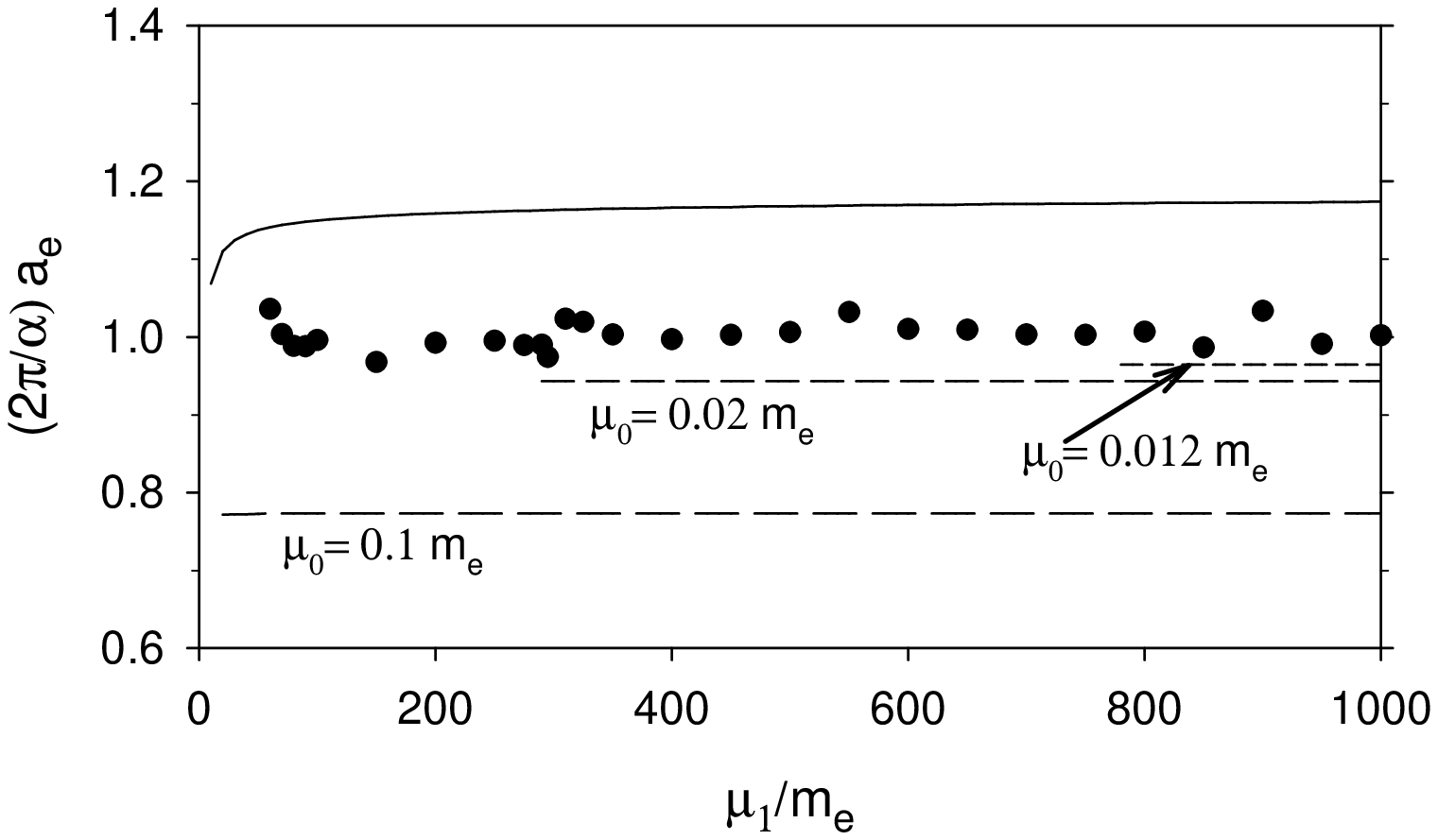}}
\caption{\label{fig:ae} The anomalous moment of the electron in
units of the Schwinger term ($\alpha/2\pi$) plotted versus
the PV photon mass, $\mu_1$.  The solid line is the 
standard-parameterization result for the one-electron/one-photon
truncation and the dashed lines are the results for
sector-dependent parameterization at three different values of
photon mass $\mu_0$; both use $m_1=\infty$.  The results for the
sector-dependent parameterization are plotted only for values
of $\mu_1$ for which the probability of the two-particle
sector remains between 0 and 1; for values of $\mu_0\leq0.01m_e$,
this requires $\mu_1>1000m_e$, which is beyond the range of
the plot.  For the case of the standard parameterization,
$\mu_0$ has its physical value of zero.
The filled circles are from a calculation with the standard parameterization
that includes the self-energy contribution from the one-electron/two-photon
sector.  It also includes a second PV photon flavor, with its mass, 
$\mu_2$, set to $\sqrt{2}\mu_1$; the PV electron mass $m_1$ 
is equal to $2\cdot10^4\,m_e$. 
The variation is due to errors in numerical quadratures.}
\end{figure}

In the limit where the PV electron mass $m_1$ is infinite,
the expression for the anomalous moment is
\bea
a_e&=&\frac{\alpha_0}{\pi}m_e^2z_0^2\int y^2 (1-y) dy dk_\perp^2 \\
  &&  \times
  \left(\sum_{k=0}^1 \frac{(-1)^k}{ym_0^2+(1-y)\mu_k^2+k_\perp^2-m_e^2y(1-y)}\right)^2 .
    \nonumber
\eea
For the sector-dependent parameterization, the product $\alpha_0 z_0^2$ is just $\alpha$,
and the bare mass $m_0$ in the denominator is replaced by the physical mass $m_e$.
To be consistent with \cite{Karmanov}, we eliminate the projection, which
does not affect the result significantly, and obtain
\bea \label{eq:secdepae}
a_e&=&\frac{\alpha}{\pi}m_e^2\int y^2 (1-y) dy dk_\perp^2 \\
  &&  \sum_{k=0}^1 (-1)^k
    \left( \frac{1}{ym_e^2+(1-y)\mu_k^2+k_\perp^2-m_e^2y(1-y)}\right)^2
    \nonumber
\eea
as the sector-dependent form of the anomalous moment in the
one-photon truncation.  In the $\mu_1\rightarrow\infty$,
$\mu_0\rightarrow0$ limit, this becomes
exactly the Schwinger result~\cite{Schwinger}, written as~\cite{BrodskyDrell}
\be \label{eq:Schwinger}
a_e=\frac{\alpha}{\pi}m_e^2\int
     \frac{ dy dq_\perp^2/(1-y)}{\left[\frac{m_e^2+q_\perp^2}{1-y}
     +\frac{q_\perp^2}{y}-m_e^2\right]^2} . 
\ee
The anomalous moment is 
infrared and ultraviolet safe, and the integral of the $k=0$ term
yields $1/2m_e^2$.  Of course, this limit cannot be taken, because the
underlying theory would become inconsistent; however, from Fig.~\ref{fig:ae},
we can infer that the $k=1$ term of (\ref{eq:secdepae})
can be small for $\mu_1$ sufficiently
large and that $\mu_0$ can be
taken small enough to attain a result near the Schwinger result.

%%%%%%%%%%%%%%%%%%%%%%%%%%%%%%%%%%%%%
%%%%%%%%%%%%%%%%%%%%%%%%%%%%%%%%%%%%%%%%%%%%%%%%%%%%%%%%%%%%%%

For the standard parameterization, $\alpha_0=\alpha$, and the
normalization is $z_0^2=1/(1+e^2J_2)$, as given in (\ref{eq:z0SR}).
The anomalous moment is then
\bea
a_e&=&\frac{\alpha m_e^2}{\pi(1+e^2 J_2)}\int y^2 (1-y) dy dk_\perp^2 \\
  &&  \times
  \left( \sum_{k=0}^1 \frac{(-1)^k}{ym_0^2+(1-y)\mu_k^2+k_\perp^2-m_e^2y(1-y)}\right)^2 .
  \nonumber
\eea
This is also plotted in Fig.~\ref{fig:ae}.  For the plotted range of $\mu_1$,
the normalization factor is unimportant.  The deviation from the Schwinger
result is due
to the presence of the bare mass in the denominator, and the fact that
the anomalous-moment integral is sensitive to constituent masses~\cite{hb}.

Although the sector-dependent approach, which inserts $m_e$ for $m_0$,
would appear to be a simple remedy, the complications already discussed
make this choice undesirable.  Instead, we can include the physics
that does adjust the mass by taking into account the self-energy
correction that comes from the one-electron/two-photon truncation~\cite{thesis},
which we consider in the next section.

We can also consider the anomalous gravitomagnetic moment, which can
be computed from a spin-flip matrix element of the energy-momentum
tensor~\cite{Brodskyetal}.  A result of zero is necessary for any
approximation to be deemed faithful to the original symmetries
of the field theory~\cite{ZeroB(0)}.  As shown in \cite{Brodskyetal},
whenever the initial and final states can be expressed as Fock-state
expansions with wave functions that depend on boost-invariant,
internal momentum variables, the contributions to
the anomalous gravitomagnetic moment are zero, term by term
in the Fock expansions.  This is precisely the case here,
for both the standard and sector-dependent parameterizations,
with (\ref{eq:projected}) as the Fock expansion to be compared with
the expansion used for $|\Psi^\updownarrow(P^+,\vec P_\perp)\rangle$
in Eq.~(59) of \cite{Brodskyetal}, 
and the net result for this anomalous moment is therefore 
zero, as required.

\section{Self-Energy Contribution}
\label{sec:selfenergy}

\subsection{Coupled Equations}

\begin{figure}[ht]
\vspace{0.1in}
\centerline{\includegraphics[width=12cm]{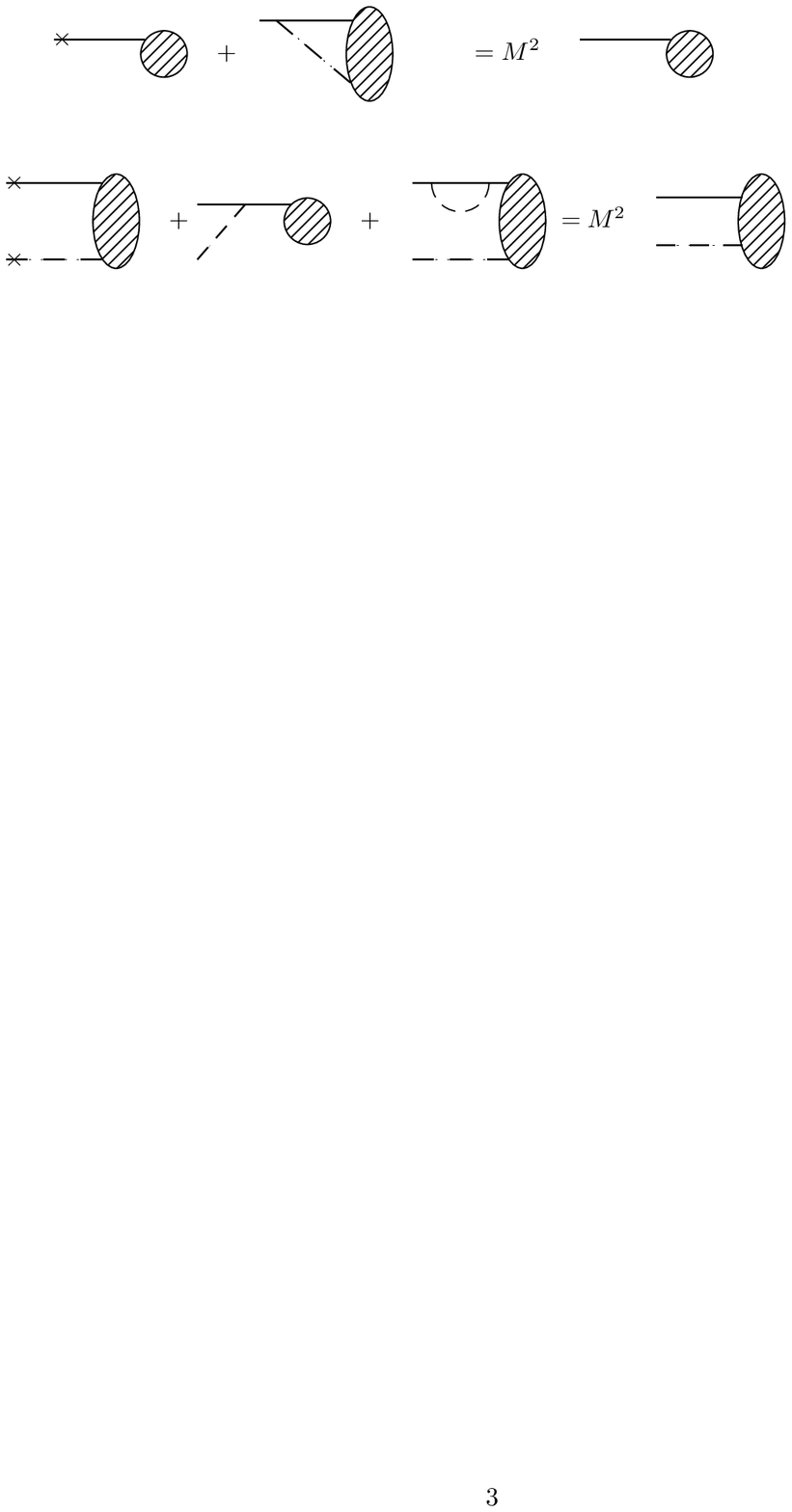}}
\caption[Diagrammatic representation of
the coupled equations.]%
{\label{fig:viscacha3} Diagrammatic representation of
the coupled equations (\ref{eq:firstcoupledequation}) and 
(\ref{eq:ReducedEqnSE}) of the text. The filled circles and ovals
represent wave functions for Fock states; the solid lines represent
fermions; and the dashed lines represent photons.  The crosses on
lines represent the light-cone kinetic energy contributions, which
are summed over all particles in the Fock state.}
\end{figure}

We extend the equations (\ref{eq:TwoBodyEqns}) and (\ref{eq:TwoBodyEqns2})
for the one-electron/one-photon
sector to include coupling to the one-electron/two-photon sector
\bea \label{eq:secondcoupledequationextended}
\lefteqn{\left[M^2 - \frac{m_i^2 + q_\perp^2}{(1-y)} - \frac{\mu_l^2 + q_\perp^2}{y}\right]
  C_{ils}^{\mu\pm}(\ub{q}) }&& \\
    &=& \sqrt{\xi_l}\sum_j (-1)^j z_j P^+ 
      \left[\delta_{s,\pm 1/2}V_{ijs}^\mu(\ub{P}-\ub{q},\ub{P})
             +\delta_{s,\mp 1/2}U_{ij,-s}^\mu(\ub{P}-\ub{q},\ub{P})\right] \nonumber \\
    && +\sum_{ab\nu}(-1)^{a+b}\epsilon^\nu\int d\ub{q}' 
          \frac{2\sqrt{\xi_b}}{\sqrt{1+\delta_{bl}\delta^{\mu\nu}}}
    \left[V_{ais}^{\nu *}(\ub{P}-\ub{q}'-\ub{q},\ub{P}-\ub{q}')
             C_{abls}^{\nu\mu\pm}(\ub{q}',\ub{q}) \right. \nonumber \\
    &&   \rule{2in}{0mm} \left.  +U_{ais}^{\nu *}(\ub{P}-\ub{q}'-\ub{q},\ub{P}-\ub{q}')
             C_{abl,-s}^{\nu\mu\pm}(\ub{q}',\ub{q}) \right] , \nonumber
\eea
and add the equation for the new sector
\bea \label{eq:thirdcoupledequation}
\lefteqn{\left[M^2 - \frac{m_i^2 + (\vec q_{1\perp}+\vec q_{2\perp})^2}{(1-y_1-y_2)} 
      - \frac{\mu_j^2 + q_{1\perp}^2}{y_1}  - \frac{\mu_l^2 + q_{2\perp}^2}{y_2}\right]
  C_{ijls}^{\mu\nu\pm}(\ub{q}_1,\ub{q}_2) }&& \\
    &=& \frac{\sqrt{1+\delta_{jl}\delta^{\mu\nu}}}{2}\sum_a (-1)^a \left\{
        \sqrt{\xi_j}\left[V_{ias}^\mu(\ub{P}-\ub{q}_1-\ub{q}_2,\ub{P}-\ub{q}_2)
                              C_{als}^{\nu\pm}(\ub{q}_2)  \right.  \right.  \nonumber \\
     && \rule{2in}{0mm}  \left.   +U_{ia,-s}^\mu(\ub{P}-\ub{q}_1-\ub{q}_2,\ub{P}-\ub{q}_2)
                              C_{al,-s}^{\nu\pm}(\ub{q}_2)\right]  \nonumber \\
      &&\rule{0.7in}{0mm}  
        + \sqrt{\xi_l}\left[V_{ias}^\nu(\ub{P}-\ub{q}_1-\ub{q}_2,\ub{P}-\ub{q}_1)
                              C_{ajs}^{\mu\pm}(\ub{q}_1)   \right. \nonumber \\
     &&  \left.\left. \rule{1.7in}{0mm}
                   +U_{ia,-s}^\nu(\ub{P}-\ub{q}_1-\ub{q}_2,\ub{P}-\ub{q}_1)
                              C_{aj,-s}^{\mu\pm}(\ub{q}_1)\right] \right\}. \nonumber
\eea
This last equation can be solved explicitly for the three-body wave function
$C_{ijls}^{\mu\nu\pm}$.  Substitution of this into 
(\ref{eq:secondcoupledequationextended}) and retention of only
self-energy contributions for the two-photon intermediate states yields
\be \label{eq:ReducedEqnSE}
\left[M^2
  -\frac{m_i^2+q_\perp^2}{1-y}-\frac{\mu_l^2+q_\perp^2}{y}\right]
C_{ils}^{\mu\pm}(y,q_\perp)=S_{ils}^{\mu\pm}
+\frac{\alpha}{2\pi}\sum_{i'}\frac{I_{ili'}(y,q_\perp)}{1-y}C_{i'ls}^{\mu\pm}(y,q_\perp),
\ee
with $i=0,1$ and $l=0,1,2$.  Here the coupling to the bare-electron sector 
is written as
\be
S_{ils}^{\mu\pm}=\sqrt{\xi_l}\sum_j (-1)^jz_j P^+
        [\delta_{s,\pm1/2} V_{ijs}^\mu(\ub{P}-\ub{q},\ub{P})
             +\delta_{s,\mp1/2} U_{ij,-s}^\mu(\ub{P}-\ub{q},\ub{P})],
\ee
and the new self-energy contributions are written in terms of the integral
\bea  \label{eq:RawSE}
I_{ili'}(y,q_\perp)
&=&(1-y)\frac{2\pi}{\alpha}  \sum_{ab\nu}\int 
    \frac{(-1)^{a+b+i'}\xi_b\epsilon^\nu d\ub{q}'}
     {M^2 - \frac{m_a^2 + (\vec q_{\perp}^{\,\prime}+\vec q_{\perp})^2}{(1-y-y')} 
      - \frac{\mu_b^2 + q_{\perp}^{\prime 2}}{y'}  
      - \frac{\mu_l^2 + q_{\perp}^2}{y} } \\
&&   \rule{0.4in}{0mm} \times \left[V_{ais}^{\nu *}(\ub{P}-\ub{q}'-\ub{q},\ub{P}-\ub{q}')     
        V_{ai's}^\nu(\ub{P}-\ub{q}'-\ub{q},\ub{P}-\ub{q})  \right.
                                \nonumber   \\     
    && \rule{0.6in}{0mm} \left. +U_{ais}^{\nu *}(\ub{P}-\ub{q}'-\ub{q},\ub{P}-\ub{q}') 
   U_{ai's}^\nu(\ub{P}-\ub{q}'-\ub{q},\ub{P}-\ub{q})\right]. \nonumber
\eea
These, combined with the coupled equation 
(\ref{eq:firstcoupledequation}) for the one-body amplitude, constitute the 
eigenvalue problem when only the self-energy contributions of the two-photon 
states are included.
A diagrammatic representation is given in Fig.~\ref{fig:viscacha3}.%
%%%%%%%%%%%%%%%%%%%%%%%%%%%%%%%%%%%%%
%%%%%%%%%%%%%%%%%%%%%%%%%%%%%%%%%%%%%%%%%%%%%%%%%%%%%%%%%%%%%%

The self-energy integral can be written in simpler form.
The change of variables $x=q^{\prime +}/q^+=y'/(1-y)$ and
$\vec{k}_\perp=\vec{q}_\perp^{\,\prime}+x\vec{q}_\perp$ yields
\be  \label{eq:selfenergy}
I_{ili'}(y,q_\perp)
=\sum_{a,b}(-1)^{i'+a+b}\xi_b\int_0^1\frac{dx}{x}\frac{d^2k_\perp}{\pi}
\frac{m_i m_{i'} -2 \frac{m_i+m_{i'}}{1-x}m_a+\frac{m_a^2+k_\perp^2}{(1-x)^2}}
{\Lambda_l-\frac{m_a^2+k_\perp^2}{1-x}-\frac{\mu_b^2+k_\perp^2}{x}} ,
\ee
which, on use of $\bar I_0$, $\bar I_1$, and $\bar J$ defined 
in Eqs.~(\ref{eq:In}) and (\ref{eq:J}), can be written as
\be
I_{ili'}(y,q_\perp)
=16\pi^2(-1)^{i'}\left[m_i m_{i'} \bar I_0(\Lambda_l)-2(m_i+m_{i'})\bar I_1(\Lambda_l)
            + \bar J(\Lambda_l)\right],
\ee
with
\be \label{eq:Lambda}
\Lambda_l\equiv \mu_l^2+(1-y)M^2-\frac{\mu_l^2+q_\perp^2}{y}.
\ee
For $\bar J$ we still have the identity 
$\bar J(\Lambda_l)=\Lambda_l \bar I_0(\Lambda_l)$.  
The integrals can be evaluated analytically,
but, for extreme values of the momentum, such as momentum fractions
on the order of $(m_0/m_1)^2\sim10^{-10}$, evaluation of the
analytic form suffers from round-off error due to the finite precision
available in floating-point calculations.  The self-energy
is then best computed by numerical evaluation of the
longitudinal integrals.  Notice
that $I_{ili'}$ includes a flavor changing self-energy, where the index $i'$ 
is different from the index $i$; this is a result of the flavor changing currents
used in the interaction Lagrangian (\ref{eq:Lagrangian}).

The solution for the coupling constant and the one-electron
amplitudes $z_i$ is
\bea
\alpha_\pm&=& \frac{G_{00}+G_{11}\pm\sqrt{(G_{00}-G_{11})^2-4G_{10}G_{01}}}
                       {16\pi[G_{00}G_{11}-G_{10}G_{01}]} ,   \\ 
\nonumber \\
\frac{z_1}{z_0}&=& \frac{[G_{11}-G_{00}]/2\mp\sqrt{(G_{00}-G_{11})^2-4G_{10}G_{01}}}{G_{01}} ,
\eea
with
\be
G_{il}=\frac{(-1)^l}{M^2-m_i^2}[m_i m_l \tilde I_0-2(m_i+m_l)\tilde I_1 +\tilde J]
\ee
and the integrals $\tilde I_n$ and $\tilde J$ defined in (\ref{eq:tildeI0I1J}).
Details of the solution are given in \ref{sec:SEsolution}.
We again have $\alpha$ as a function of $m_0$ and the PV masses,
and we seek the value of $m_0$ that yields the physical value of $\alpha$.
The calculation requires numerical quadrature of the integrals
$\tilde I_n$ and $\tilde J$; this can be done with the quadrature schemes
discussed in \cite{YukawaTwoBoson} and \cite{thesis}.

\subsection{Anomalous Moment}

The wave functions $C_{ijs}^{\mu\pm}$ can be used to compute the
anomalous moment.  Since $\tilde J$ does not satisfy any known
identity, it cannot be eliminated. Therefore, the calculation must
be done at finite $m_1$, and the second PV photon flavor must be
kept~\cite{ChiralLimit}.  However, the calculation can still 
be done~\cite{thesis}.  

The results are included in Fig.~\ref{fig:ae}.
The value of the PV electron mass $m_1$ is chosen to be $2\cdot10^4\,m_e$,
which was found in the case of the one-photon truncation to be
sufficiently large~\cite{ChiralLimit}.  The ratio of PV photon masses $\mu_2/\mu_1$
is held fixed at $\sqrt{2}$, and $\mu_1$ is varied.  
As for the choice of $m_1$, this choice was found sufficient in \cite{ChiralLimit}.
The results are consistent with perturbative QED, showing only variations
expected from numerical errors of order 1\% in calculating the underlying
integrals $\tilde I_n$ and $\tilde J$.

That the self-energy contribution brings the result so close to the leading
Schwinger contribution is clear from the following.
The dominant contribution to the expression (\ref{eq:OnePhotonae})
for the anomalous moment is the $j=0$, $i=i'=0$, and $k=k'=0$ 
contribution to the first term;
the other terms are suppressed by the large PV masses that 
appear in the denominators of the wave functions $C_{iks}^{\mu\pm}(\ub{k})$.
For the dominant term, the denominator, as determined 
by (\ref{eq:SEwavefunctions}) and (\ref{eq:Aij}), is essentially the square of
\be
A_{00}=\frac{m_0^2+\frac{\alpha}{2\pi}I_{000}+q_\perp^2}{1-y}
        +\frac{\mu_0^2+q_\perp^2}{y}-m_e^2,
\ee
with
\be
I_{000}=16\pi^2[(m_0^2+\Lambda_0)\bar I_0(\Lambda_0)-4m_0\bar I_1(\Lambda_0)]
\ee
and
\be
\Lambda_0=\mu_0^2+(1-y)m_e^2-\frac{\mu_0^2+q_\perp^2}{y},
\ee
from the expressions in (\ref{eq:selfenergy}) and (\ref{eq:Lambda}).
For the physical photon, $\mu_0$ is zero, and the two-body wave function
is peaked at $q_\perp=0$ and $y=0$, so that we can approximate
$\Lambda_0$ as $m_e^2$.  We then have
\be
A_{00}\simeq
  \frac{m_0^2+8\pi\alpha[(m_0^2+m_e^2)\bar I_0(m_e^2)-4m_0\bar I_1(m_e^2)]+q_\perp^2}{1-y}
  +\frac{q_\perp^2}{y}-m_e^2
\ee
In our formulation, the perturbative one-loop electron self-energy
can be read from Eq.~(\ref{eq:FeynEigen}) for $i=0$, with
$z_1=0$, $M^2=m_0^2+\delta m^2$ on the left, and $M^2=m_0^2$
on the right.  This yields
\be \label{eq:oneloopSE}
\delta m^2=m_0^2-m_e^2
   =2e^2\left[m_0^2\bar I_0(m_0^2)-4m_0\bar I_1(m_0^2)+\bar J(m_0^2)\right].
\ee
From this mass shift, we have, to leading order in $\alpha=e^2/4\pi$,
with use of the identity $\bar J(m_0^2)=m_0^2\bar I_0(m_0^2)$,
\be
m_0^2=m_e^2-8\pi\alpha[(m_0^2+m_e^2)\bar I_0(m_e^2)-4m_0\bar I_1(m_e^2)].
\ee
Therefore, the denominator reduces to the square of
\be
A_{00}=\frac{m_e^2+q_\perp^2}{1-y}+\frac{q_\perp^2}{y}-m_e^2+{\cal O}(\alpha^2),
\ee
which matches the denominator of the integral (\ref{eq:Schwinger})
that yields the Schwinger contribution.
Thus, the dominant contribution to the anomalous moment
with the self-energy included is essentially the same as
the integral that yields the Schwinger result.

%%%%%%%%%%%%%%%%%%%%%%
\section{Summary}
\label{sec:summary}
%%%%%%%%%%%%%%%%%%%%%%

In the calculation presented here, we have continued the development
of nonperturbative Pauli--Villars regularization and light-front
Hamiltonian techniques as a method for the determination of
bound-state wave functions in quantum field theories.  The new
results extend previous work on QED~\cite{OnePhotonQED,ChiralLimit}
to include some of the effects of one-electron/two-photon Fock
states; the previous work was limited to a one-electron/one-photon
truncation.  To reach agreement with the Schwinger
result for the anomalous moment~\cite{Schwinger} is meant as a test
of the method but not as the purpose of the method.  Instead, the method
is intended for strongly coupled theories, where perturbation theory
is not applicable.  Applications of the method to a gauge theory are
important as precursors to applications to QCD.

We have also shown that use of a sector-dependent 
parameterization~\cite{SectorDependent,hb,Karmanov}
requires great care in the handling of an infrared divergence
and an uncanceled ultraviolet divergence.  
The expression for the anomalous moment is itself safe from
these divergences, and one is led to think that regulators
can be removed.  However, the underlying theory is not safe
and quickly enters an unphysical regime, if care is not taken.
The bare coupling becomes imaginary, the Fock-state wave 
functions become ill-defined, and the Fock-sector probabilities
fall outside the interval $[0,1]$.

The infrared and ultraviolet divergences are interconnected.
The infrared divergence is regulated by the introduction
of a nonzero mass for the physical photon, but the
result for the anomalous moment of the electron
does not agree with experiment unless this mass is quite small.
In turn, a small physical photon mass requires a large PV photon
mass to keep Fock-sector probabilities within the physical
range of zero to one.
Unfortunately, a large PV photon mass can lead
to great difficulties for a numerical
calculation with a higher-order Fock truncation,
as was seen in \cite{thesis}.

The standard parameterization does not have an infrared
problem and has no particular difficulty with its own
uncanceled ultraviolet divergence.  The Fock-state
wave functions are well defined.
The one-electron/one-photon truncation does yield~\cite{ChiralLimit}
an anomalous moment that is 17\% larger than the experimental
value; in a sense, it includes too much physics from 
higher orders in $\alpha$.  The discrepancy with 
experiment, which is unrelated to the uncanceled divergence,
is immediately corrected by the self-energy
contribution from the one-electron/two-photon sector,
which adjusts the constituent electron mass that appears
in the anomalous moment integral.
The sector-dependent approach, in effect, 
attempts to incorporate this correction
by forcing the constituent mass to be equal to the
physical mass, which causes the infrared divergence.
Thus, the standard approach is to be 
preferred, at least for gauge theories.

%%%%%%%%%%%%%%%%%%%%%%
\section*{Acknowledgments}
%%%%%%%%%%%%%%%%%%%%%%
This work was supported in part by the Department of Energy
through Contract No.\ DE-FG02-98ER41087.

\appendix

%%%%%%%%%%%%%%%%%%%%%%%%%%%%%%%%%%%%%%%%%
\section{Feynman-gauge QED}  \label{sec:FeynmanQED}
%%%%%%%%%%%%%%%%%%%%%%%%%%%%%%%%%%%%%%%%%%%

The PV-regulated Feynman-gauge QED Lagrangian is
\bea \label{eq:Lagrangian}
{\cal L} &= & \sum_{i=0}^2 (-1)^i \left[-\frac14 F_i^{\mu \nu} F_{i,\mu \nu} 
         +\frac12 \mu_i^2 A_i^\mu A_{i\mu} 
         -\frac{1}{2} \left(\partial^\mu A_{i\mu}\right)^2\right] \\
  && + \sum_{i=0}^1 (-1)^i \bar{\psi_i} (i \gamma^\mu \partial_\mu - m_i) \psi_i 
  - e_0 \bar{\psi}\gamma^\mu \psi A_\mu ,  \nonumber
\eea
where
\begin{equation} \label{eq:NullFields}
  A_\mu  = \sum_{i=0}^2 \sqrt{\xi_i}A_{i\mu}, \;\;
  \psi =  \sum_{i=0}^1 \psi_i, \;\;
  F_{i\mu \nu} = \partial_\mu A_{i\nu}-\partial_\nu A_{i\mu} .
\end{equation}
The subscript $i=0$ denotes a physical field and $i=1$ or 2 a PV
field.  Fields with odd index $i$ are chosen to be negatively
normed.  The constants $\xi_i$ satisfy the following
constraints~\cite{ChiralLimit}:
\bea
   \label{eq:1stconstraint}
&&\xi_0=1, \\
   \label{eq:2ndconstraint}
&&\sum_{i=0}^2(-1)^i\xi_i=0, \\
   \label{eq:3rdconstraint}
&&\sum_{i=0}^2(-1)^i\xi_i
   \frac{\mu_i^2/m_1^2}{1-\mu_i^2/m_1^2}\ln(\mu_i^2/m_1^2)=0 . 
\eea
The second constraint guarantees the necessary cancellations
for ultraviolet regularization; it also implies that $A^\mu$
in (\ref{eq:NullFields}) is a zero-norm field.  
The third guarantees the correct chiral limit
at one loop; for truncations that include higher loops, there are
order-$\alpha$ corrections to the constraint.
Implementation of the gauge condition $\partial^\mu A_{i\mu}=0$
is discussed in \cite{ChiralLimit}.

When the PV electron mass is sufficiently large, the third constraint 
(\ref{eq:3rdconstraint})
can be approximated by\footnote{In \protect\cite{ChiralLimit},
the approximation to this third constraint and the solution for $\xi_2$
are written incorrectly, without the factors of $m_1$ and without correct
subscripts for the PV photon masses, $\mu_1$ and $\mu_2$.}
\be \label{eq:chiralconstraint}
\sum_l(-1)^l\xi_l\mu_l^2\ln(\mu_l/m_1)=0. 
\ee
The solution to the set of constraints, (\ref{eq:1stconstraint}), 
(\ref{eq:2ndconstraint}), and (\ref{eq:chiralconstraint}),
assuming $\mu_0=0$, is then
\be
\xi_1=1+\xi_2 \;\;  \mbox{and} \;\; 
\xi_2=\frac{\mu_1^2\ln(\mu_1/m_1)}
     {\mu_2^2\ln(\mu_2/m_1)-\mu_1^2\ln(\mu_1/m_1)}.
\ee
Without loss of generality, we require $\mu_2>\mu_1$, so that $\xi_2$
is positive.

The dynamical fields are
\bea
\psi_{i+}&=&\frac{1}{\sqrt{16\pi^3}}\sum_s\int d\ub{k} \chi_s
  \left[b_{is}(\ub{k})e^{-i\ub{k}\cdot\ub{x}}
        +d_{i,-s}^\dagger(\ub{k})e^{i\ub{k}\cdot\ub{x}}\right]\,, \\
A_{i\mu}&=&\frac{1}{\sqrt{16\pi^3}}\int \frac{d\ub{k}}{\sqrt{k^+}}
  \left[a_{i\mu}(\ub{k})e^{-i\ub{k}\cdot\ub{x}}
        +a_{i\mu}^\dagger(\ub{k})e^{i\ub{k}\cdot\ub{x}}\right]\,,
\eea
with~\cite{LepageBrodsky} $\chi_s$ an eigenspinor of $\Lambda_+\equiv\gamma^0\gamma^+/2$.
The creation and annihilation operators satisfy (anti)commutation
relations
\bea
\{b_{is}(\ub{k}),b_{i's'}^\dagger(\ub{k}'\}
   &=&(-1)^i\delta_{ii'}\delta_{ss'}\delta(\ub{k}-\ub{k}'), \\
\{d_{is}(\ub{k}),d_{i's'}^\dagger(\ub{k}'\}
   &=&(-1)^i\delta_{ii'}\delta_{ss'}\delta(\ub{k}-\ub{k}'), \\
{[}a_{i\mu}(\ub{k}),a_{i'\nu}^\dagger(\ub{k}']
   &=&(-1)^i\delta_{ii'}\epsilon^\mu\delta_{\mu\nu}\delta(\ub{k}-\ub{k}').
\eea
Here $\epsilon^\mu = (-1,1,1,1)$ is the metric signature for the
photon field components in Gupta--Bleuler quantization~\cite{GuptaBleuler,GaugeCondition}.

Without antifermion terms, the Hamiltonian is
\begin{eqnarray} \label{eq:QEDP-}
\lefteqn{{\cal P}^-=
   \sum_{i,s}\int d\ub{p}
      \frac{m_i^2+p_\perp^2}{p^+}(-1)^i
          b_{i,s}^\dagger(\ub{p}) b_{i,s}(\ub{p})} \\
   && +\sum_{l,\mu}\int d\ub{k}
          \frac{\mu_l^2+k_\perp^2}{k^+}(-1)^l\epsilon^\mu
             a_{l\mu}^\dagger(\ub{k}) a_{l\mu}(\ub{k})
          \nonumber \\
   && +\sum_{i,j,l,s,\mu}\int d\ub{p} d\ub{q}\left\{
      b_{i,s}^\dagger(\ub{p}) \left[ b_{j,s}(\ub{q})
       V^\mu_{ij,2s}(\ub{p},\ub{q})\right.\right.\nonumber \\
      &&\left.\left.\rule{0.5in}{0in}
+ b_{j,-s}(\ub{q})
      U^\mu_{ij,-2s}(\ub{p},\ub{q})\right] 
            \sqrt{\xi_l}a_{l\mu}^\dagger(\ub{q}-\ub{p})
                    + H.c.\right\} .  \nonumber
\end{eqnarray}
The vertex functions are~\cite{OnePhotonQED}
\begin{eqnarray} \label{eq:vertices}
    V^0_{ij\pm}(\ub{p},\ub{q}) &=& \frac{e_0}{\sqrt{16 \pi^3 }}
                   \frac{ \vec{p}_\perp\cdot\vec{q}_\perp
                      \pm i\vec{p}_\perp\times\vec{q}_\perp
                       + m_i m_j + p^+q^+}{p^+q^+\sqrt{q^+-p^+}} , \\
    V^3_{ij\pm}(\ub{p},\ub{q}) &=& \frac{-e_0}{\sqrt{16 \pi^3}}
                        \frac{ \vec{p}_\perp\cdot\vec{q}_\perp
                      \pm i\vec{p}_\perp\times\vec{q}_\perp
                       + m_i m_j - p^+q^+ }{p^+q^+\sqrt{q^+-p^+}} , \nonumber\\
    V^1_{ij\pm}(\ub{p},\ub{q}) &=& \frac{e_0}{\sqrt{16 \pi^3}}
       \frac{ p^+(q^1\pm i q^2)+q^+(p^1\mp ip^2)}{p^+q^+\sqrt{q^+-p^+}} , \nonumber\\
    V^2_{ij\pm}(\ub{p},\ub{q}) &=& \frac{e_0}{\sqrt{16 \pi^3}}
       \frac{ p^+(q^2\mp i q^1)+q^+(p^2\pm ip^1)}{p^+q^+\sqrt{q^+-p^+}} , \nonumber\\
    U^0_{ij\pm}(\ub{p},\ub{q}) &=& \frac{\mp e_0}{\sqrt{16 \pi^3}}
       \frac{m_j(p^1\pm ip^2)-m_i(q^1\pm iq^2)}{p^+q^+\sqrt{q^+-p^+}} , \nonumber\\
    U^3_{ij\pm}(\ub{p},\ub{q}) &=& \frac{\pm e_0}{\sqrt{16 \pi^3}}
       \frac{m_j(p^1\pm ip^2)-m_i(q^1\pm iq^2)}{p^+q^+\sqrt{q^+-p^+}} , \nonumber\\
    U^1_{ij\pm}(\ub{p},\ub{q}) &=& \frac{\pm e_0}{\sqrt{16 \pi^3}}
                            \frac{m_iq^+-m_jp^+ }{p^+q^+\sqrt{q^+-p^+}} , \nonumber\\
    U^2_{ij\pm}(\ub{p},\ub{q}) &=& \frac{i e_0}{\sqrt{16 \pi^3}}
                     \frac{m_iq^+-m_jp^+ }{p^+q^+\sqrt{q^+-p^+}} . \nonumber
\end{eqnarray}
For the standard approach, $m_0$ is the same for all sectors and $e_0$ is 
just $e$, there being no fermion-antifermion loops included in the calculation.
For the sector-dependent approach, $m_0$ and $e_0$ depend on the Fock sector
where the Hamiltonian is applied.  The sector-dependent constants could more
generally be functions of momentum~\cite{SectorDependent,hb}, but here we
follow \cite{Karmanov} and use them as constants.

\section{Solution for the Self-Energy Contribution}  \label{sec:SEsolution}

The two-body integral equations (\ref{eq:ReducedEqnSE}) can be expressed compactly as
\bea  \label{eq:SEMixingEquations}
A_{0j}C_{0js}^{\mu\pm} - B_jC_{1js}^{\mu\pm} &=& -S_{0js}^{\mu\pm} \\
B_j C_{0js}^{\mu\pm} + A_{1j} C_{1js}^{\mu\pm} &=& -S_{1js}^{\mu\pm}. \nonumber
\eea
where $A_{ij}$ and $B_j$ are defined by
\be  \label{eq:Aij}
A_{ij}=\frac{m_i^2+q_\perp^2}{1-y}+\frac{\mu_j^2+q_\perp^2}{y}
    +\frac{\alpha}{2\pi}\frac{I_{iji}}{1-y}-M^2
\ee
and
\be \label{eq:Bj}
B_j=\frac{\alpha}{2\pi}\frac{I_{1j0}}{1-y}=-\frac{\alpha}{2\pi}\frac{I_{0j1}}{1-y}.
\ee
We solve this 2$\times$2 system for the two-body wave functions
\be \label{eq:SEwavefunctions}
C_{ijs}^{\mu\pm}=-\frac{A_{1-i,j}S_{ijs}^{\mu\pm}+(-1)^iB_j S_{1-i,js}^{\mu\pm}}
                        {A_{0j}A_{1j}+B_j^2}.
\ee
Without the self-energy contributions, we have $B_j=0$, and the wave functions
reduce to the forms given in (\ref{eq:wavefn1}) and (\ref{eq:wavefn2}),
with their line of poles whenever $m_0<m_e$, as discussed in 
Sec.~\ref{sec:onephoton}.  Here, however, the self-energy contributions
make the denominators more complicated.  For values of $m_0$ that are smaller
than $m_e$ by an amount of order $\alpha$, there need not be a line
of poles.  In fact, we find that, for the solution with self-energy
contributions, there is no pole in this two-body Fock sector.

Substitution of (\ref{eq:SEwavefunctions}) into (\ref{eq:firstcoupledequation}),
and use of the expressions (\ref{eq:vertices}) for the vertex functions, yields
\be \label{eq:FeynEigenSE}
[M^2-m_i^2]z_i=2e^2\sum_j(-1)^j z_j [m_i m_j \tilde I_0
                                   -2(m_i+m_j)\tilde I_1 +\tilde J],
\ee
where
\bea \label{eq:tildeI0I1J}
\tilde I_0&=&\int \frac{dy dq_\perp^2}{16\pi^2}\sum_j (-1)^j\xi_j
      \frac{A_{0j}-A_{1j}-2B_j}{y[A_{0j}A_{1j}+B_j^2]}, \\
\tilde  I_1&=&\int \frac{dy dq_\perp^2}{16\pi^2}\sum_j (-1)^j\xi_j
   \frac{m_1A_{0j}-m_0A_{1j}-(m_0+m_1)B_j}{y(1-y)[A_{0j}A_{1j}+B_j^2]}, \nonumber \\
\tilde  J&=&\int \frac{dy dq_\perp^2}{16\pi^2}\sum_j(-1)^j\xi_j \nonumber \\
&& \times
  \frac{(m_1^2+q_\perp^2)A_{0j}-(m_0^2+q_\perp^2)A_{1j}-2(m_0m_1+q_\perp^2)B_j}
                  {y(1-y)^2[A_{0j}A_{1j}+B_j^2]} .  \nonumber
\eea
When the self-energy contributions are neglected, these return to the previous
expressions Eq.~(\ref{eq:In}) and (\ref{eq:J}) for $\bar I_0$, $\bar I_1$, and 
$\bar J$ in the one-photon truncation.  What is more, the eigenvalue equation 
for $z_i$ has nearly the same form as the eigenvalue equation (\ref{eq:FeynEigen}) 
in the one-photon case.  The only difference in finding the analytic solution is 
that $\tilde I_0$ and $\tilde J$ are not connected by any known identity.

%%%%%%%%%%%%%%%%%%%%%%%%%%%%%%%%
\bibliographystyle{model1a-num-names}

\end{document}